\documentclass[12pt,letterpaper]{article}
\usepackage{epsfig,rotating,setspace,latexsym,amsmath,epsf,amssymb,amsfonts,bm,theorem,cite,caption,subcaption,enumerate,longtable,accents,bbm,appendix}
\usepackage{algorithm,algorithmic,graphicx,epsf,authblk,epstopdf,url,color,multirow}
\usepackage{mathtools,comment,tabularx}
\usepackage{comment}
\usepackage[inline]{enumitem}

\usepackage{dsfont}

\newtheorem{theorem}{Theorem}

\newtheorem{corollary}{Corollary}

\newtheorem{proposition}{Proposition}
\newtheorem{remark}{Remark}
\newtheorem{observation}{Observation}
\newtheorem{lemma}{Lemma}

\newenvironment{Proof}[1]{\medskip\par\noindent{\bf Proof:\,}\,#1}{{\mbox{\,$\blacksquare$}\par}}

\newcolumntype{Y}{>{\centering\arraybackslash}X}
\newcommand{\figref}[1]{\figurename~\ref{#1}}
\allowdisplaybreaks

\title{Transaction Capacity, Security and Latency in Blockchains}

\author{Mustafa Doger \qquad Sennur Ulukus\\
\normalsize Department of Electrical and Computer Engineering\\
\normalsize University of Maryland, College Park, MD 20742\\
\normalsize  \emph{doger@umd.edu} \qquad \emph{ulukus@umd.edu}}

\begin{document}
\date{}
\maketitle

\begin{abstract}
We analyze how secure a block is after the block becomes $k$-deep, i.e., security-latency, for Nakamoto consensus under an exponential network delay model. We provide the fault tolerance and extensive bounds on safety violation probabilities given mining rate, delay rate and confirmation rules. Next, modeling the blockchain system as a batch service queue with exponential network delay, we connect the security-latency analysis to sustainable transaction rate of the queue system. As our model assumes exponential network delay, batch service queue models give a meaningful trade-off between transaction capacity, security and latency. Our results indicate that, by simply picking $k=7$-block confirmation rule in Bitcoin instead of the convention of $k=6$, mining rate, latency and throughput can be increased sixfold with the same safety guarantees. We further consider adversarial attacks on the queue service to hamper the service process. In an extreme scenario, we consider the selfish-mining attack for this purpose and provide the maximum adversarial block ratio in the longest chain under the exponential delay model. The ratio in turn reflects the maximum rate of decrease in the sustainable transaction rate of the queue.
\end{abstract}

\section{Introduction}\label{sec::intro}
Blockchains allow users to maintain a distributed ledger of accounts. Nakamoto's white paper \cite{btc-whitepaper} described a mining procedure based on proof of work (PoW) and a longest chain protocol that enables users to agree on a single version of a ledger. This protocol, now widely known as Nakamoto consensus, requires  users to mine a new block at the tip of their longest chain of blocks. Each block in this chain contains a nonce that proves the validity of the block and rewards the miner who created it compensating for its mining costs. 

In blockchain systems based on Nakamoto consensus, some blocks are mined on the same height in the chain due to network delays and create forks that split the network into multiple versions of the ledger. In addition to these honest forks caused by network conditions, adversarial activities can create deliberate forks as well. Forks in turn, undermine the consensus among the users by disrupting agreement on a single version of the ledger. As mining process continues, the forks resolve in favor of one of the competing ledgers. Hence, blocks are not confirmed immediately to prevent changes in the set of confirmed set of blocks and transactions inside. The more one waits to confirm a block, the safer the content of the block will be, i.e., it is less likely that another block at the same height (another version of the ledger) is confirmed after the block is confirmed. This is referred to as the \emph{security-latency problem}, and has been extensively analyzed by various different models and protocols \cite{garay, non-lockstep-sec-lat, nakamoto-always-wins, guo-close-sec-lat, guo-btc-sec-lat, our-sec-lat-isit, our-sec-lat-extended, gazi-sec-lat}. Under bounded network delay $\Delta$, \cite{guo-close-sec-lat} analyzes how secure a block is after waiting $t$ time units whereas \cite{guo-btc-sec-lat, our-sec-lat-extended} answer the same security question using block generations as time units.

As transactions arrive to the blockchain system, they are included in the next blocks that are to be mined based on the fee they offer to the miner. The miner of a block can only include up to a certain number of transactions or the block can have up to a certain size of storage units. In this sense, the blockchain systems can be modeled as a priority-queue, where transactions enter a queue and compete to be included in the next block. Different batch service queues have been modeled for blockchain analysis \cite{Kasahara-2019,malak-queue,q-li-queue-info-theory,ricci-experimental-queue-model}. \cite{Kasahara-2019} has given an insightful batch service queue model and has evaluated the average service time and correctly interpreted that the micro-payments are not sustainable for the Bitcoin system. \cite{malak-queue} models a batch service queue for Bitcoin that gives an estimate service time of a transaction given the fee and current state of the mempool. Since each mined block experiences a network delay until the other miners can observe it, which decreases the rate of service of transactions, \cite{quan-li-queue-blockchain} re-models the batch service queue system that accounts for the network delay and gives the stability condition for the queue system as well as a numerical method to find the steady state of the system.

Although state-of-the-art methods focus on bounded delay models to analyze the security in blockchains, recently, \cite{pow-under-random-safe} showed that PoW longest chain protocols have good security guarantees even when delays are sporadically large and possibly unbounded. In this paper, we first give bounds on the security-latency question, i.e., how secure a block is after a certain number of blocks are published, where we re-model the network delay with exponential distribution and consider a rigged model first introduced in \cite{guo-btc-sec-lat}. This analysis in turn is interpreted as how fast blocks can be mined given honest  fraction of the hash power in the system, a confirmation rule and a security threshold. Then, we show that the security level can be parameterized in terms of the proportion of mining rate to the network delay which we call $\kappa$, also known as fork rate in the literature \cite{information_propagation}. Next, the security-latency analysis is connected to the queue model via the stability condition of the batch service queue system which enables us to reinterpret the number of transactions a blockchain system can sustain as a function of the block size, network conditions and fork rate $\kappa$. We also consider several queue service attacks and their effect on sustainable transaction rate. Although we consider a first-come, first-served queue model, this model could be interpreted as a queue system, where we only consider transactions in the service that have a high priority, i.e., it is a first-come, first-served queue model for transactions that pay a sufficient fee. Otherwise, it does not make sense to consider the stability condition as the low-fee transactions are too many to have a stable system in any case. This priority queue model is inspired by the two-class queue model of \cite{Kasahara-2019} and can also be interpreted as that of \cite{malak-queue}, where the transactions in the queue are considered as pending only when they pay more fee than the fee a user is willing to pay.

Our findings imply that for given network conditions, a block size $b$, honest fraction of the total hash power $\alpha$, safety violation probability threshold $\overline{p}$, increasing $k$-block confirmation rule allows the mining rate to increase which in turn increases the sustainable transaction rate. This adjustment of confirmation rule and mining rate first drops the expected transaction confirmation time until a point where a minimum is achieved and further increasing $k$-block confirmation rule starts to increase the expected transaction confirmation time, i.e., a trade-off emerges.

For current Bitcoin parameters, i.e., $300$ seconds of block interarrival time and $6$-block confirmation rule, using our model, under $10\%$ adversarial presence, the upper bound on the transaction safety violation probability is found to be $0.00118$ and $\approx7$ transactions per second is shown to be sustainable. Using the same block size and network conditions for Bitcoin with the same safety violation probability threshold, by picking $7$-block confirmation rule and increasing the mining rate so that a block is mined every $\approx90$ seconds, sustainable transaction rate increases sevenfold to $\approx49$, and a transaction is confirmed after $\approx10$ minutes instead of current confirmation latency of one hour. By keeping $10^{-3}$ as safety violation probability threshold and switching $6$-block confirmation rule to $8$-block confirmation rule, a block can be mined every $\approx30$ seconds, which increases sustainable transaction rate up to $\approx140$ and achieves transaction confirmation latency of $\approx4$ minutes. Under adversarial attacks on the queue-service process, we show that throughput decreases at most $\approx11\%$($\approx33\%$ and $\approx67\%$ resp.) under $10\%$ ($25\%$ and $40\%$ resp.) adversarial presence.

Another recent work \cite{Kiffer_bounded_capacity_security} studies the block production capacity in PoW and PoS longest chain protocols where the authors consider adversarial attacks on the block processing capacity of honest miners. The study builds on \cite{neu_bandwith_constraint_security} and models a queue where honest miners need to process the block headers and the adversary strategically times the release of its blocks to cause congestion in this queue. In this paper, we do not consider queue attacks on block headers as in \cite{Kiffer_bounded_capacity_security}. On the other hand, different than \cite{Kiffer_bounded_capacity_security}, we consider adversarial attacks on the state of the mempool queue where adversarial blocks override the honest blocks that revert the progress in the mempool queue service. An early work on the security and performance of blockchains \cite{on_the_sec_pow_gervais} studies parameter changes in Bitcoin network and its effect on selfish-mining and double-spending with an MDP model and simulations. Our work in comparison, is an explicit, in detail analysis of double-spending and queuing that provides theoretical overview of the parameter changes and its implications on capacity and security-latency. We also note that rather than changing the parameters, some works focus on changing the Nakamoto consensus protocol to achieve better throughput \cite{bitcoin_ng,PHANTOM_GHOSTDAG,SPECTRE,FruitChains,Sompolinsky2015SecureHT}. 

Concurrent with our work, we became aware of \cite{cao2023tradeoff}, which established a new model to analyze the security-latency trade-off for bounded delay model that works for safety parameters up to ultimate fault tolerance and replaces previously established rigged model of \cite{guo-btc-sec-lat, our-sec-lat-isit} with a partial Markov chain that counts the number of successful balanced attacks by the adversary. The results are connected to throughput by simply multiplying the block size with the rate of growth under $\Delta$-delay strategy explained in \cite{our-sec-lat-isit}. An optimization problem is formulated for the throughput and security-latency and a trade-off is established between latency and throughput for a given security level. Different from \cite{cao2023tradeoff}, we independently used an exponential delay model which is then connected to the throughput via a batch-service queue system and its stability condition. As the stability condition for exponential delay model, first established in \cite{quan-li-queue-blockchain}, ends up with the same form as the product of the block size with the rate of growth under $\Delta$-delay strategy for bounded delay model, optimistic results for the throughput in both methods agree with each other. In addition to optimistic throughput analysis, here, we consider two different queue-attack strategies, a mild one and a pessimistic one in order to reflect the change in throughput under different situations. Although an analogous throughput-latency result can be established with an optimization problem via our analysis, since we argue against an optimization problem that treats block size independent from security level, our interpretation of the trade-off differs from \cite{cao2023tradeoff}, where we pick a security level and a block size then compare different $k$-confirmation rules and mining rates. 

\section{System Model}
In this paper, we assume a blockchain protocol that employs PoW and longest chain rules, that works as follows. The system consists of $n$ miner nodes who compete for mining the next block where $n$ is large. A block essentially consists of a pointer, a nonce and a ledger. The pointer points to the previous block which it extends. The ledger of accounts inside the block extends the ledger history of the chain of blocks that can be reached via the pointer. To make a block valid, in addition to having correct semantic details, e.g., valid transactions in the ledger, the miners need to find a valid nonce. In this context, the process of finding a valid nonce is called \textit{mining}. The semantic details regarding the validity of transactions and block contents together with a valid nonce can be abstracted out as a Poisson process with mining (arrival) rate of $\mu_2$. Note that mining can be represented as performing Bernoulli trials until a valid nonce is found. Since $n$ is large and each miner can try multiple nonces each second, the mining process converges to a Poisson process as mentioned. This mining process is essentially conducted by $n$ (assumed to be large enough) miner nodes in the system each of which having infinitesimal hash (mining) power. Out of these $n$ nodes, $\alpha$ fraction are honest and they follow the longest chain protocol, i.e., they mine on the tip of the longest chain they have seen so far. The other nodes, which make up $\beta=1-\alpha$ fraction, are adversarial and are allowed to deviate from the longest chain protocol as long as the block they mine contains a valid nonce.
 
We further assume that a block contains at most $b$ transactions all of which have to be semantically correct. The transactions arrive with an exponential rate of $\lambda$ to the mempools of the nodes to be included in the next blocks and are served based on first-come, first-served principle. We assume a transaction arrives to all mempools at the same time (each honest node has its own mempool based on the longest chain they observed). Additionally, each block at height $h$ that is mined by an honest node is subject to a network delay of $t_{hi}\in [0,\Delta_h]$ before it becomes available to any other honest node $i$, where $\Delta_h \sim Exp(\mu_1)$ where $\Delta_h$ are i.i.d. In other words, once an honest node mines a block $b_h$ at height $h$ with a valid nonce, the block becomes available to other honest nodes after some delay which is at most $\Delta_h \sim Exp(\mu_1)$. In this model, a single adversary is assumed to be controlling all $\beta$ mining power (fully-coordinating adversaries) and the network delay as long as $t_{hi}\in [0,\Delta_h]$, and ties are broken in adversary's favor. A transaction in this model is said to be confirmed according to a $k$-block confirmation rule if it is part of the longest chain of an honest view and $k$-deep, i.e., there are at least $k-1$ blocks mined on top of it. We denote this system model as $\mathbbm{B}(\alpha,b,\mu_1,\mu_2,\lambda,k)$.

It is important to point out the difference of the delay model here from those that are commonly used for the security-latency analysis in the literature. Here, we assume an honest block at height $h$ is delayed for some $t_{hi}\in [0,\Delta_h]$ where $\Delta_h \sim Exp(\mu_1)$ before it is available to honest node $i$ whereas delays are taken to be bounded between zero and some constant maximum delay $\Delta$ in \cite{gazi-sec-lat,guo-btc-sec-lat,guo-close-sec-lat,nakamoto-always-wins}, i.e., $t_{hi}\in[0,\Delta]$ for $\forall h,i$. Models with constant maximum delays were proven to be secure against adversarial behavior with $\beta<\frac{1-\beta}{1+(1-\beta)\lambda\Delta}$ in the long-run \cite{nakamoto-always-wins}. We provide our conjecture for the ultimate fault tolerance for exponential delay model in Appendix~\ref{sec::app-c}. Further, in Section~\ref{sec::sec-lat analysis}, we give loose conditions under which the model is secure and rigorously calculate an upper bound for safety violation probability, and show that this probability decays with $k$.

Further, we note that a slightly different version of this system model without any adversarial behavior and maximum exponential delay, i.e., $t_{hi}=\Delta_h$ for all $i$ was studied as a two-step batch service queue in \cite{quan-li-queue-blockchain}. Our main goal here is to bring latency-security analysis and queue analysis studied in the literature together. More specifically, we want to find the maximum transaction arrival rate $\lambda$ a blockchain system can sustain under different adversarial attacks, given honest fraction $\alpha$, block size $b$, safety violation threshold $p$. Table~\ref{table::freq-table} shows some frequently used variables throughout this paper to ease the navigation of the paper for the reader. 

\begin{table}[t]
    \begin{center}
        \begin{tabular}{||c c||} 
             \hline
             Parameters & Definitions \\ [0.5ex] 
             \hline\hline
             $\lambda$ & transaction arrival rate \\
             \hline
             $\alpha$ & honest fraction of all nodes (power)   \\ 
             \hline
             $\beta$ & $1-\alpha$ \\
             \hline
             $\mu_1$ & rate of network delay \\
             \hline
             $\mu_2$ & rate of mining  \\
             \hline
             $\sigma$ & $\frac{\mu_1}{\mu_1+\mu_2}$  \\
             \hline
             $\rho$ & $1-\sigma$\\
             \hline
              $\sigma'$ & $\frac{\mu_1}{\mu_1+\mu_2\beta}$  \\
             \hline
             $\rho'$ & $1-\sigma'$\\
             \hline
             $\Bar{a}_i$ & $\alpha\rho^i+\beta\cdot\mathbbm{1}_{i\leq2}$\\
             \hline
              &  \\   [-1.1em]
             $\Bar{b}_i$ & $\alpha\rho^i+\beta\cdot\mathbbm{1}_{i\leq1}$\\
             \hline
             $\kappa$ & $\frac{\rho}{\sigma}=\frac{\mu_2}{\mu_1}$\\
             \hline
        \end{tabular}
    \end{center}
    \caption{Frequently used notations.}
    \label{table::freq-table}
\end{table}

\section{Rigged Private Attack} \label{sec::sec-lat analysis}
Our goal in this section is to give an upper bound on the safety violation probability. To do so, we consider a rigged jumper model with private attack for adversarial strategy, to be defined next and denoted as $\overline{\mathbbm{B}}(\alpha,b,\mu_1,\mu_2,\lambda,k)$. Consider the genesis block, which is at height $h=0$ and available to all nodes at time $t=0$ and called the zeroth jumper block. At time $t>0$, the honest node $i$ is trying to mine a new block on top of the genesis block that contains $\min(b,|Q_i(t)|)$ transactions, where $Q_i(t)$ is the mempool at time $t$ observed by honest node $i$. The block $b_1$, which is the first honest block mined on top of the genesis block contains the least number of transactions compared to any other honest block at the same height $h=1$. This is due to the fact that, for each miner $i$, $|Q_i(t)|$ is monotonically increasing until a new block enters the view of the honest miner $i$. Hence, among the blocks at $h=1$, the miner of the first block observes the lowest $|Q_i(t)|$. In this rigged model, after this first honest block on $h=1$ is mined, say at time $t_1$, all other honest blocks mined by other honest nodes on the same height $h=1$ between $t_1$ and $t_1+\Delta_1$ are converted to an adversarial block, i.e., they are rigged. As we consider an upper bound on the safety violation, we assume that the adversary can append the rigged blocks on a different height than the height that honest miner originally created it. For example, when the adversary has a private chain, it can append the rigged block to the tip of its private chain and treat it as a valid adversarial block. Note that, the adversary cannot append the block on the tip of a block that is not mined yet. As a result, $b_1$, which we call the first jumper, that was mined first on $h=1$, will enter the view of all other honest nodes at time $t_1+\Delta_1$ and will be the only block at this height for all of them. Note that, we assume that no more than one block can be mined at exactly the same time since this event has zero probability. As $n$ is large, each miner has infinitesimal hash power, hence the miner of the first block cannot create a block at height $h=2$ during ${(}t_1,t_1+\Delta_1]$. Thus, after $t_1+\Delta_1$, all miners try to mine a new block on $h=2$ and we assume the same conversion of honest blocks for $h=2,\ldots$, i.e., after an honest miner creates a block at each height $h$, the block observes maximum $\Delta_h$ delay and all other blocks are rigged. We call this mining model as the rigged jumper model. Note that in this model, we have a single honest block on each height. It is proven in \cite{guo-btc-sec-lat} that \emph{private attack} is the best attack to cancel a transaction after it is confirmed under the condition that \underline{a}ll \underline{h}onest \underline{b}locks are \underline{o}n \underline{d}ifferent \underline{h}eights (we denote this condition in short as AHBODH) for bounded $\Delta$-delay model. Next, we restate this theorem proven in \cite{guo-btc-sec-lat} for exponential network delay. We skip proving this theorem as it essentially follows the same steps as bounded $\Delta$-delay model, by simply replacing $\Delta$ with $\Delta_h \sim Exp(\mu_1)$ in \cite[Lemma~8]{guo-btc-sec-lat} and  \cite[Corollary~11]{guo-btc-sec-lat}.

\begin{theorem}
    {\normalfont \textbf{(Guo-Ren\cite{guo-btc-sec-lat})}} Under AHBODH, if any attack succeeds in violating a transaction's safety then the private mining attack also succeeds in violating that transaction's safety.
    \label{thm::ahbodh}
\end{theorem}

The safety violation probability, $\overline{p}(\alpha,\mu_1,\mu_2,k)$ of $\overline{\mathbbm{B}}(\alpha,b,\mu_1,\mu_2,\lambda,k)$ is an upper bound on the safety violation probability $p(\alpha,\mu_1,\mu_2,k)$ of $\mathbbm{B}(\alpha,b,\mu_1,\mu_2,\lambda,k)$ which follows from Theorem~\ref{thm::ahbodh} and the fact that the adversary is given the advantage of rigged blocks. In other words, assuming $t_{hi}=\Delta_h$ for all honest $i$, i.e., each honest node $i$ experiences maximum possible delay for any honest block at height $h$ and rigged adversarial blocks make sure that the adversary is strictly more powerful than it actually is and AHBODH is satisfied, thus we can upper bound the probability of violating a transaction's safety. Note that, in the model described as $\overline{\mathbbm{B}}(\alpha,b,\mu_1,\mu_2,\lambda,k)$, all mempools during $[t_h+\Delta_h,t_{h+1})$ are the same. Further, as any honest block mined during $(t_h,t_h+\Delta_h]$ is rigged, the mempools observed by honest nodes are effectively the same all the time, which is denoted as $Q(t)$. We call the mining process of building blocks with a valid nonce, controlled by rate $\mu_2$, as the block-generation process. The network delay process after a block is mined, which is controlled by rate $\mu_1$, is called blockchain-building process.

\subsection{Security-Latency Analysis}
We are interested in security-latency analysis of a certain high priority transaction $tx$ which arrives at the mempool at time $\tau$, i.e., we would like to find the safety violation probability which is the probability that $tx$ is discarded after it is confirmed under the $k$-block confirmation rule. Here, for simplicity, we assume at the time of arrival of $tx$, $|Q(\tau)|< b$. The rigorous analysis done in this section can be extended for the case where $|Q(\tau)|\geq b$, which is explained in Appendix~\ref{sec::app-b}. As mentioned earlier, the safety violation probability of $\mathbbm{B}(\alpha,b,\mu_1,\mu_2,\lambda,k)$ is upper bounded by the safety violation probability of $\overline{\mathbbm{B}}(\alpha,b,\mu_1,\mu_2,\lambda,k)$. Hence, we use the rigged private attack to find $\overline{p}(\alpha,\mu_1,\mu_2,k)$, an upper bound for $p(\alpha,\mu_1,\mu_2,k)$.

\subsection{Pre-Mining Gain}
We start by analyzing the lead of the adversary at time $\tau$, which is the difference between the longest chain and the longest honest chain just before the transaction arrives at the system. As it was done in \cite{our-sec-lat-extended,guo-btc-sec-lat,guo-close-sec-lat}, we assume that $\tau$ is large, thus we model the lead as the stationary distribution of an extended birth-death process of the rigged model. The steady state distribution of the following extended birth-death Markov chain transition matrix will be equal to adversary's lead:
\begin{align}
    P=\begin{bmatrix}
        \alpha \sigma & \alpha\sigma\rho+\beta & \alpha\sigma\rho^2  & \alpha\sigma\rho^3 & \alpha\sigma\rho^4 & \ldots\\
        \alpha \sigma & \alpha\sigma\rho & \alpha\sigma\rho^2+\beta  & \alpha\sigma\rho^3 & \alpha\sigma\rho^4 & \ldots\\
        0          & \alpha \sigma & \alpha\sigma\rho & \alpha\sigma\rho^2+\beta  & \alpha\sigma\rho^3 & \ldots\\
        0          & 0 & \alpha \sigma & \alpha\sigma\rho & \alpha\sigma\rho^2+\beta   &  \ldots\\
        \vdots & \vdots & \vdots & \vdots & \vdots & \ddots \\
    \end{bmatrix}. \label{eq::pmatrix}
\end{align}
To explain the reasoning, assume that the blockchain-building process is not active at some random time, i.e., no block with a valid nonce is experiencing network delay (e.g., right after genesis-block is published at $t=0$). At such a time, we define $T_i$ as a new ``mining event'' $i$, i.e., the block-generation process creates a new block. Let $A_i$ denote the event that this block is adversarial and let $H_i$ denote the event that the block is an honest jumper. If $T_i=H_i$, i.e., if the newly created block at ``mining event'' $i$ is honest jumper, let $D_{i,j}$ denote the event that the total number of blocks created during the network delay of this honest jumper block equals $j$. Note that, 
\begin{align}
    T_i&=A_i\cup (\bigcup_{j=0}^{\infty}H_i\cap D_{i,j}),
\end{align}
and
\begin{align}
    P(T_i=A_i)&=\beta,\\
    P(T_i=H_i)&=\alpha,\\
    P(D_{i,j}|T_i=H_i)&=\rho^j\sigma.
\end{align}

The events defined above are sufficient to explain $P$ where each state $l$ represents the adversarial lead $l$. Assume that adversarial lead is $l>0$ and the blockchain-building process is not active at some random time. Hence, the block-generation process will create a new ``mining event'' $T_i$:
\begin{enumerate}
    \item With probability $\beta$, $T_i=A_i$ and the adversary does not share the block, hence we do not need to consider network delay in this case. Thus, the lead increases by $1$, i.e., the Markov chain transitions from $l$ to $l+1$. (Note, after this point, the blockchain-building process is not active and the block-generation process will create a new ``mining event'' again.)
    \item With probability $\alpha$, $T_i=H_i$ and we consider the network delay on the blockchain-building process following this ``mining event.'' Before the honest block is published, i.e., during the network delay which is distributed exponentially with $\mu_1$, we consider how many additional blocks can be mined. Honest blocks will be converted to adversarial during this delay time due to the rigged model. Since the mining times are distributed exponentially with rate $\mu_2$, $D_{i,j}$ happens with probability $\sigma\rho^j$ and all $j$ blocks will extend the adversarial lead (honest blocks will be rigged). Hence, the lead increases from $l$ to $l+j-1$ by the time the honest jumper block is published with probability $\sigma\rho^j$.
\end{enumerate}

After each of these two cases, the blockchain-building process is not active and the block-generation process will create a new ``mining event'' again. Hence, each $T_i$ is independent from each other. Note that the lead cannot go below $0$ since when adversary has no lead, it can mine its private chain on top of the newly mined honest block, hence we have a birth-death process. The first row is accounting for this fact as we assume that the adversary is aware of any mined block immediately (upper bound). In other words, in the first row, the lead increases from $l$ to $l+j$ by the time the honest jumper block is published with probability $\sigma\rho^j$. 

Note that the $P$ in \cite{our-sec-lat-extended} is structurally the same as here (i.e., a stochastic matrix of M/G/1 type) even though the delay models are different. Next, we find the steady state of matrix $P$ using Ramaswami's formula \cite{ramaswami-mg1}.
\begin{lemma} \label{lemma::ramaswami}
    {\normalfont \textbf{(Ramaswami\cite{ramaswami-mg1})}} Steady state of $P$ can be found recursively using 
    \begin{align}
        \pi_i&=\frac{\pi_0 \Bar{b}_i+\sum_{j=1}^{i-1}\pi_j\Bar{a}_{i+1-j}}{1-\Bar{a}_1}, \quad i\geq1, 
        \label{eq::ramaswami-form}
    \end{align}
    where
    $\pi_0=3-\sigma^{-1}-\alpha^{-1}$.
\end{lemma}
\begin{Proof}
    Note that, finding $\Bar{a}_i$ and $\Bar{b}_i$ is a straightforward application of Ramaswami's formula and they are provided in Table~\ref{table::freq-table}. Here, we only prove that $\pi_0=3-\sigma^{-1}-\alpha^{-1}$.
    
    Summing \eqref{eq::ramaswami-form} for $i\geq1$, we get,
    \begin{align}
        (1-\Bar{a}_1)\sum_{i\geq1}\pi_i&=\sum_{i\geq1}\left(\pi_0 \Bar{b}_i+\sum_{j=1}^{i-1}\pi_j\Bar{a}_{i+1-j}\right)\\
        &=\pi_0\sum_{i\geq1}\Bar{b}_i+\sum_{i\geq2}\sum_{j=1}^{i-1}\pi_j\Bar{a}_{i+1-j}\\
        &=\pi_0\left(\alpha\frac{\rho}{\sigma}+\beta\right)+\sum_{j\geq1}\sum_{i>j}\pi_j\Bar{a}_{i+1-j}\\
        &=\pi_0\left(\alpha\frac{\rho}{\sigma}+\beta\right)+\sum_{j\geq1}\pi_j\sum_{i\geq2}\Bar{a}_{i}\\
        &=\pi_0\left(\alpha\frac{\rho}{\sigma}+\beta\right)+\sum_{i\geq1}\pi_i\left(\alpha\frac{\rho^2}{\sigma}+\beta\right).
    \end{align}
    Substituting $\sum_{i\geq1}\pi_i=1-\pi_0$ and  $(1-\Bar{a}_1)=\alpha\sigma$ and expanding terms,
    \begin{align}
        \alpha\sigma(1-\pi_0)&=\pi_0\left(\alpha\frac{\rho^2}{\sigma}+\alpha\frac{\rho\sigma}{\sigma}+\beta\right)+(1-\pi_0)\left(\alpha\frac{\rho^2}{\sigma}+\beta\right)\\
        &=\pi_0\alpha\rho+\alpha\frac{\rho^2}{\sigma}+\beta.
    \end{align}
    Finally,
    \begin{align}
        \pi_0&=\sigma-\frac{\rho^2}{\sigma}-\frac{\beta}{\alpha}\\
        &=\frac{\sigma^2-\rho^2}{\sigma}-\frac{\beta}{\alpha}\\
        &=1-\frac{\rho}{\sigma}-\frac{\beta}{\alpha}
    \end{align}
    gives the desired result.
\end{Proof}

We denote the distribution of the lead as $\Pi$ where $P(\Pi=i)=\pi_i$.

\subsection{Confirmation Interval} 
At time $\tau$, which is assumed to be large, target $tx$ enters the system and will be included in the next honest block by assumption (e.g., high priority). Note that, $\tau$ may correspond to a time of the network delay immediately following a jumper that arrived before $\tau$. In this case, due to the rigged model, if an honest block containing $tx$ is mined during this network delay, it forks the last jumper and becomes void and rigged. For example, if the last jumper before $\tau$ arrived at $\tau-\frac{\Delta}{2}$, the mining process for a block that contains $tx$ effectively starts at $\tau+\frac{\Delta}{2}$. Thus, $tx$ is effectively included in the honest chain after the end of the network delay immediately following the last jumper of the pre-mining phase. In this sense, when a $tx$ enters mempool at time $t_{tx}$ where the mining time of the last jumper block before $t_{tx}$ is denoted as $t_{tx}^-$ (and its delay as $\Delta^-$), we simply assume $\tau=\max(t_{tx},t_{tx}^-+\Delta^-)$. In other words, without loss of generality, $tx$ enters the mempool at a time $\tau$, where the next honest block to be mined contains the $tx$ and is a jumper. Notice that, our bound for the lead in previous section considers the leads at time instances corresponding to the end of the delay intervals of jumper blocks, i.e., $t_i+\Delta_i$ (for example $t_{tx}^-+\Delta^-$), hence the assumption is consistent with the results of the previous section. 

The duration starting with $\tau$ until the time that $tx$ is confirmed by the honest miners, i.e., the block containing $tx$ becomes $k$-deep in all honest views, is called the confirmation interval. Here, we note that, from the perspective of safety violation, we should consider a transaction confirmed if it is part of the longest chain in any honest view, since honest miners are not sure about each others views. However, as the blocks that cause forks are rigged and each miner has infinitesimal hash power, in this model, the views are essentially the same and do not change during the delay intervals. Hence, for the ease of the analysis, we consider `all honest views'. Further, as it is pointed out in \cite[Section III.B]{our-sec-lat-extended}, statistically, the bounds on the violation probability do not change by this choice. Denoting the mining time of the $k$th jumper block after $\tau$ as $\tau_k$ and exponential delay of the $k$th jumper block as $\Delta_k$, confirmation interval spans $[\tau,\tau_k+\Delta_k]$.

Our goal is to find how many adversarial (including rigged) blocks are mined privately during the confirmation interval. To do that, we first find the distribution of the number of adversarial blocks mined between the publication times of two honest jumpers, denoted $C$.

\begin{lemma}\label{lemma::lower-geo}
The number of adversarial (including rigged) blocks mined between the publication times of two consecutive jumper blocks $C$ has the following distribution,
\begin{align}
    P_{C}(c)=\alpha\beta^{c}\sigma \sum_{j=0}^{c}\left(\frac{\rho}{\beta}\right)^j.
    \label{eq::geo-C}
\end{align}
\end{lemma}

\begin{Proof}
 Assume a mining event $T'_0$ that results in an honest jumper, i.e., $T'_0=H'_0$ and the jumper is followed by some number of blocks mined at its network delay. After this honest (jumper) is published, let $T'_i$ denote the $i$th next ``mining event''. Notice,
\begin{align}
    P(\inf\{i\geq1:T'_i=H'_i\}=m+1)=\beta^m\alpha,
\end{align}
i.e., there can be $m$ adversarial arrivals before the arrival of the next jumper with probability $\beta^m\alpha$. With the arrival of this honest block, we also have,
\begin{align}
    P(D_{m+1,j}|\inf\{i\geq1:T'_i=H'_i\}=m+1)=\rho^j\sigma,
\end{align}
i.e., there can be additional $j$ block arrivals (including the rigged blocks) with probability $\rho^j\sigma$ before the honest block is published under exponential network delay. 
Hence, we have,
\begin{align}
    P_{C}(c)=\sum_{j=0}^{c}P(\inf\{i\geq1:T'_i=H'_i\}=c-j+1)\cdot P(D_{c-j+1,j}|\inf\{i\geq1:T'_i=H'_i\}=c-j+1),
\end{align}
which results in \eqref{eq::geo-C}. This jumper block will be published at the end of the exponential network delay. Moreover, by the memoryless property, the next jumper also has the same distribution and they are independent, thus, they are i.i.d.
\end{Proof}

The confirmation interval contains $k$ jumper blocks, hence the distribution of the number of adversarial blocks mined privately during the confirmation interval, denoted as $S_k$, is equal to the sum of $k$ i.i.d.~random variables $C$.

\begin{corollary}\label{cor::conf}
    The number of adversarial (including rigged) arrivals during the confirmation interval, denoted as $S_k$, has the following distribution,
\end{corollary}
\begin{align}
    P_{S_k}(s) &= \alpha^k\sigma^k\beta^s\sum_{n=0}^{s}\binom{k\!-\!1\!+\!n}{n}\binom{k\!-\!1\!+\!s\!-\!n}{\!s\!-\!n}\left(\frac{\rho}{\beta}\right)^n. \label{eq::pascal-like}
\end{align}

\begin{Proof}
    We start by computing the probability generating function (PGF) of $C$,
    \begin{align}
        G_C(z)&=\alpha\sigma\sum_{c\geq0}(z\beta)^c\sum_{j=0}^{c}\left(\frac{\rho}{\beta}\right)^j\\
        &=\alpha\sigma\sum_{j\geq0}\sum_{c\geq j}(z\beta)^c\left(\frac{\rho}{\beta}\right)^j\\
        &=\alpha\sigma\sum_{j\geq0}\frac{(z\rho)^j}{1-z\beta}\\
        &=\frac{\alpha\sigma}{(1-z\beta)(1-z\rho)}.
    \end{align}
    Next, PGF of $S_k$ denoted as $G_{S_k}(z)$, is,
    \begin{align}
         G_{S_k}(z)&= G_C(z)^k\\
         &=\alpha^k\sigma^k\frac{1}{(1-z\beta)^k}\frac{1}{(1-z\rho)^k}\\
         &=\alpha^k\sigma^k\sum_{m\geq0}\binom{k\!-\!1\!+\!m}{m}z^m\beta^m\sum_{n\geq0}\binom{k\!-\!1\!+\!n}{n}z^n\rho^n\\
         &=\sum_{s\geq0}\alpha^k\sigma^k z^s\beta^s\sum_{n=0}^{s}\binom{k\!-\!1\!+\!n}{n}\binom{k\!-\!1\!+\!s\!-\!n}{\!s\!-\!n}\left(\frac{\rho}{\beta}\right)^n,
    \end{align}
    which completes the proof.
\end{Proof}

\subsection{Post-Confirmation Race}
The block containing target transaction $tx$ will be confirmed when $k$-deep at $\tau_k+\Delta_k$. At the same time, a conflicting block with transaction $tx'$ at the adversarial chain will be $(\Pi+S_k)$-deep. If $\Pi+S_k\geq k$, then the adversary can publish its private chain and cancel $tx$. Else, the race enters the post-confirmation phase, where the adversarial deficit is $D=k-\Pi-S_k$, i.e., the difference between the length of the longest honest chain containing $tx$ and the longest adversarial chain containing $tx'$. In this part of the race, the adversary has to make up for its deficit $D$, in order to discard the confirmed transaction $tx$. If at any time after $\tau_k+\Delta_k$, the adversary's private chain is able to catch the longest honest chain containing $tx$, it can publish its chain to cancel $tx$. Next, let $M=\max\limits_{i\geq1} (S'_i-i+1)$ and $S'_i=\sum_{j'=1}^{i}C'_{j}$, where $C'_{j}$ are  i.i.d.~with $C'_{j}\sim C$.
\begin{lemma}\label{lemma::equality_defcit_steady}
    $M=\max\limits_{i\geq1} (S'_i-i+1)$ is identically distributed as $\Pi$.
\end{lemma}
\begin{Proof}
    First note that $M=C'_1+max_{i\geq0}\sum_{j=1}^{i}(C'_{j+1}-1)$. Next, let
    \begin{align}
        \max_{i\geq0}\sum_{j=1}^{i}(C'_{j}-1)=\Tilde{M},
    \end{align}
    and 
    \begin{align}
        M_{i+1}=\max(M_i+C_i-1,0).
    \end{align}
    Using \cite[Corollary 3.2]{asmussen-ruin}, $M_{i+1}$ converges in distribution to $\Tilde{M}$, i.e.,
    \begin{align}
        \lim_{i\rightarrow\infty}P(M_{i+1}= n)=P(\Tilde{M} = n)=\Tilde{m}_n,
    \end{align}
    where $\Tilde{m}_0=\frac{1-\mathbb{E}[C]}{P(C=0)}$, and
    \begin{align}
        \Tilde{m}_i=\frac{\sum_{j=0}^{i-1}\Tilde{m}_j P(C>i-j)}{P(C=0)}.
    \end{align} 
    Next, reapplying Ramaswami's formula, it is straightforward to show that $\Tilde{M}+C$ is identically distributed as $\Pi$.
\end{Proof}

\begin{theorem}\label{lemma::rigorous-lemma-formula}
    Given mining rate $\mu_2$, honest fraction $\alpha$, delay rate $\mu_1$ and confirmation depth $k$, a confirmed transaction cannot be discarded with probability greater than,
    \begin{align}
         \overline{p}(\alpha,\mu_1,\mu_2,k)&=P(\max\limits_{i\geq1} (S'_i-i+1)\geq D)\\ &= P(\max\limits_{i\geq1} (S'_i-i+1)\geq k-\Pi-S_k)\\
    &=P(\Pi+S_k+\Pi'\geq k)\label{eq::safety-vio}
    \end{align}
    where $\Pi'$ is  i.i.d.~with $\Pi'\sim\Pi$.
\end{theorem}

\begin{Proof}
Note that, $\max\limits_{i\geq1} (S'_i-i+1)$ is nonnegative since for $i=1$, $S'_1=C'_1\geq0$. Hence, $D\leq0$, which denotes the fact that the adversary has at least $k$ blocks on top of $tx'$ when $tx$ is $k$-deep, is counted as immediate double-spending as we want. Consider $D>0$ and beginning from $\tau_k+\Delta_k$, let us denote $\tau_{k+h}$ as $\tau'_{h}$. Once the honest miners mine an honest jumper block $b'_{1}$ at $\tau'_{1}$, which will be published at $\tau'_{1}+\Delta'_{1}$, there will be $C'_{1}\sim C$ adversarial blocks mined during $[\tau_k+\Delta_k,\tau'_{1}+\Delta'_{1}]$. If $C'_{1}\geq D$, the adversary can publish its private chain right before $\tau'_{1}+\Delta'_{1}$, discarding the transaction from some honest view. This explains the case where $S'_1\geq D$. Assume $C'_{1}< D$, consider the total number of adversarial blocks mined from $\tau_k+\Delta_k$ until the second honest jumper's publication time at $\tau'_{2}+\Delta'_{2}$, i.e., during $[\tau_k+\Delta_k,\tau'_{2}+\Delta'_{2}]$, which is equal to $S'_2=C'_{1}+C'_{2}$. Right before $\tau'_{2}+\Delta'_{2}$, $tx$ is $k+1$-deep at honest miners views (except the miner of $b'_{2}$), whereas $tx'$ is $\Pi+S_k+S'_2$-deep. If $\Pi+S_k+S'_2 \geq k+1$, $tx$ will be discarded from some honest view, which explains $S'_2-1 \geq D$. In general, for $i>0$, right before $\tau'_{i}+\Delta'_{i}$, $tx$ is $k+i-1$-deep at honest miners views (except the miner of $b'_{i}$), whereas $tx'$ is $\Pi+S_k+S'_i$-deep, hence we need $S'_i-i+1 \geq D$ for some $i$.
\end{Proof}

\begin{corollary}\label{cor::fault-tolerance}
{\normalfont \textbf{(Fault tolerance)}} $\lim_{k \to \infty}\overline{p}(\alpha,\mu_1,\mu_2,k)\rightarrow0$ iff
\begin{align}
    \alpha>\frac{1}{(2-\kappa)}.\label{eq::fault-tolerance}
\end{align}
\end{corollary}

\begin{Proof}
      Note that from Lemma \ref{lemma::ramaswami}, replacing $\sigma^{-1}$ with $1+\kappa$, \eqref{eq::fault-tolerance} is necessary and sufficient condition for positive recurrence of the birth-death chain of $P$. Thus, the random walk escapes to infinity for $\alpha\leq\frac{1}{(2-\kappa)}$, i.e., the adversary is able to build a lead of more than $k$, almost surely. Else, i.e., for $\alpha>\frac{1}{(2-\kappa)}$, in the steady state distribution, the walk is at a finite value with probability $1$. Further, the maximum adversarial advantage in post-confirmation race is shown to be identically distributed as the steady state distribution of $P$, hence, it has a finite value as well. During $[\tau,\tau_k+\Delta_k]$, the walk can take negative values as well, but this does not change the dynamics of the walk and the transition probabilities. Thus, we can simply check if $\lim_{k \to \infty}P(\Pi\geq k)$. 
\end{Proof}

\begin{proposition}\label{prop::relationship_mu_s}
    $\overline{p}(\alpha,\mu_1,\mu_2,k)=\overline{p}(\alpha,1,\kappa,k)$.
\end{proposition}

\begin{Proof}
    First, note that $\beta=1-\alpha$, $\rho=1-\sigma$. Hence, the distributions of $\Pi$, $S_k$ and $M\sim\Pi$ are solely expressible in terms of $\alpha$, $\sigma$ and $k$ as a result of Lemmas~\ref{lemma::ramaswami} and \ref{lemma::lower-geo}.
    Further, $\sigma$ and $\kappa$ have one-to-one relationship given in Table \ref{table::freq-table}. Thus, replacing $(\mu_1,\mu_2)$ with $(1,\kappa)$ does not change $\overline{p}(\alpha,\mu_1,\mu_2,k)=P(\Pi+\Pi'+S_k\geq k)$.
\end{Proof}

Proposition~\ref{prop::relationship_mu_s} essentially means that the relationship between $\mu_1$ and $\mu_2$ has to be kept constant to keep $\overline{p}(\alpha,b,\mu_1,\mu_2,k)$ at a certain level, hence, $\kappa$ is a safety parameter,  which is also known as fork rate \cite{information_propagation}, and investigated in similar manners in \cite{Sompolinsky2015SecureHT,nakamoto-always-wins,cao2023tradeoff}. In other words, given $\alpha$, $k$ and a safety violation threshold $\overline{p}$, the maximum $\kappa$ possible, denoted as $\overline{\kappa}$, can be found from the analysis provided here, i.e., $\overline{\kappa}=f(\alpha,k,\overline{p})$ for some function $f$. We note that finding $f$ explicitly is a complex task, which we avoid here since our main goal is to establish a relationship between $\mu_1,\mu_2$ and security, which we already did in Proposition~\ref{prop::relationship_mu_s}.

There is an extensive discussion about choosing parameters of a blockchains in \cite{guo-close-sec-lat} when designing a blockchain while taking security-latency into account, e.g., choosing the mining difficulty $\mu_2$ in relation to $\Delta$. Our security-latency analysis in this section implies that one can consider $\mu_2$ and $\mu_1$ as dependent parameters when designing a blockchain with a certain security level. Moreover, network delay $\mu_1$ is necessarily a function of the block size $b$, i.e., $\mu_1=g(b)$ for some function $g$, which depends on the network \cite{sublinear-prop-delay}. Thus, given $b$, $\alpha$, $k$ and a safety violation threshold $\overline{p}$, one can choose the maximum mining rate as,
\begin{align}
    \overline{\mu}_2&=\overline{\kappa}\mu_1\\
    &=f(\alpha,k,\overline{p})g(b). \label{eq::mining-rate-breakdown}
\end{align}
Further, it is shown in \cite{sublinear-prop-delay}, that the block propagation delay (here $\frac{1}{\mu_1}$) increases sublinearly with block size $b$ and the real-world data for $g(b)$ resembles a piecewise linear function.

Note that the security of the content of a block decreases with the increasing mining rate. As a result, the mining rate should decrease with the increasing block size $b$ to keep the security of the transactions in a block at a certain level since a valid nonce protects the whole block together with all its transactions. The proportional relationship of $g(b)$ and $\mu_2$ shown in \eqref{eq::mining-rate-breakdown} is a side effect of the network delay that depends on the block size $b$, however, it does not consider the security of all $b$ transactions in a single block since $\overline{p}$ analyzed in this section considers the probability that a transaction is discarded. However, the adversary can potentially discard more than one transaction (up to $b$ transactions if they happen to be in the same block) with the same attack which is successful with probability less than $\overline{p}$. In this sense, a designer should consider $\overline{p}$ as a security threshold for $b$ transactions while determining $\mu_2$. Finding $f$ rigorously is a complex task as mentioned previously. However, one can use bisection method on \eqref{eq::safety-vio} to find the maximum possible $\overline{\kappa}$ which satisfies $\overline{p}(\alpha,1,\overline{\kappa},k)<\overline{p}$.

\begin{remark}
    \eqref{eq::safety-vio} gives an upper bound for the safety violation probability. For a lower bound on $p(\alpha,\mu_1,\mu_2,k)$, the distributions of $\Pi$ and $S$ have to be replaced as follows: A lower bound for the lead can be found by replacing $\sigma$ and $\rho$ with $\sigma'$ and $\rho'$ in Lemma~\ref{lemma::ramaswami} (replacing $\sigma$ and $\rho$ in definitions of $\Bar{a}_i$ and $\Bar{b}_i$ as well). A lower bound for the adversarial arrivals during the confirmation interval can be found by replacing $\sigma$ and $\rho$ with $\sigma'$ and $\rho'$ in \eqref{eq::pascal-like}. A lower bound on the maximum deficit adversary can recover can be shown to be equal to the lower bound on the lead using the same idea as in Lemma~\ref{lemma::equality_defcit_steady}.
\end{remark}

\section{Queue Analysis}\label{sec::queue}
\begin{figure}[t]
	\centerline{\includegraphics[width=0.9\columnwidth]{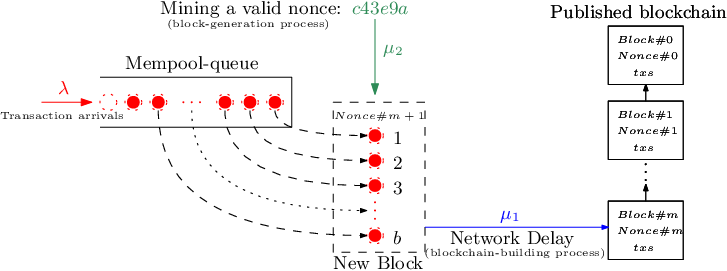}}
	\caption{Queue model.}
	\label{fig::queue}
\end{figure}

To consider the maximum rate $\lambda$ the blockchain can sustain under our system model, we resort to the queue analysis in \cite{quan-li-queue-blockchain}. \figref{fig::queue} describes the Markovian queue process of \cite{quan-li-queue-blockchain}. Transactions arrive at the mempool-queue, $Q(t)$, with a rate of $\lambda$, waiting to be included in the blockchain on a first-come, first-served basis. All honest nodes observe the same mempool-queue at any time $t$, i.e., $Q(t)$. In the first stage of the service, called block-generation process, the nodes try to mine a block with a valid nonce, which has rate $\mu_2$. When a valid nonce is found at any time $t$, this block will contain the first $\min (b,|Q(t)|)$ transactions from $Q(t)$. After the block is formed with a valid nonce, this block will be added to the blockchain according to the blockchain-building process, i.e., it will experience a network delay with rate $\mu_1$. It is assumed that no other new block is created until the newly mined block is shared with other miners. The steady state behavior of this Markovian queue process was given in an open form with an iterative formula. Here we only restate its state transition relation and main stability conditions. We refer the interested reader to \cite{quan-li-queue-blockchain,quan-li-gen-queue-blockchain}.

The state of the system at time $t$ is taken as $w(t)=(|B(t)|,|Q(t)|)$, where $|B(t)|$ and $|Q(t)|$ are the number of transactions in the new block and the queue, respectively. Note that $w(t)\in \Omega$, where
\begin{align}
    \Omega=\{(i,j): i=0,1,\ldots,b, \quad j=0,1,\ldots\}.
\end{align}

\begin{figure}[t]
	\centerline{\includegraphics[width=0.9\columnwidth]{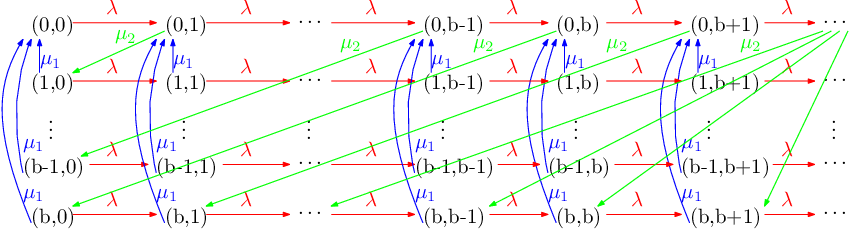}}
	\caption{State transition relation.}
	\label{fig::st-transition-q}
\end{figure}

\figref{fig::st-transition-q} describes the state transition relationship. The transition rate matrix of this system has the $GI/M/1$ form, as given below,
\begin{align}
    \mathbf{Q} = \begin{bmatrix}
        B_0 & A_0 & & & & & & &\\
        B_1 & A_1 &A_0& & & & & &\\
        B_2 &  & A_1 &A_0 & & & & &\\
        \vdots &  & & \ddots&\ddots & & & &\\
        B_b &  & & & A_1 &A_0& & &\\
        & A_b  & & & & A_1 &A_0 & &\\
         &  & A_b & & & & A_1 &A_0 &\\
         &  & & \ddots & & & & \ddots&\ddots\\
    \end{bmatrix},
\end{align}
where
\begin{align}
    A_{0}=\lambda I, \ A_{1}=\begin{bmatrix}
        -(\lambda+\mu_2) &0 &\ldots &0\\
         \mu_1&-(\lambda+\mu_1)&\ldots &0\\
        \mu_1&0 &\ddots &0\\
        \mu_1&0&\ldots  &-(\lambda+\mu_1)\\
    \end{bmatrix}, \
    A_b=\begin{bmatrix}
        0 &\ldots&0 &\mu_2\\
        0 &\ldots&0 &0\\
        \vdots &\vdots&\vdots &\vdots\\
        0 &\ldots&0 &0\\
    \end{bmatrix}
\end{align}
and
\begin{align}
    B_{0}=\begin{bmatrix}
        -\lambda &0 &\ldots &0\\
         \mu_1&-(\lambda+\mu_1)&\ldots &0\\
        \mu_1&0 &\ddots &0\\
        \mu_1&0&\ldots  &-(\lambda+\mu_1)\\
    \end{bmatrix}, \ B_{1}=\begin{bmatrix}
        0 &\mu_2&0&\ldots&0\\
        0 &0&0&\ldots&0\\
        \vdots &\vdots&\vdots &\vdots&\vdots\\
        0 &0&0 &0&0\\
    \end{bmatrix}, \
    B_{i}=\begin{bmatrix}
        \ldots&0&\mu_2&0&\ldots\\
        \ldots &0&0&0&\ldots\\
        \vdots &\vdots&\vdots&\vdots&\vdots \\
        \ldots &0&0&0&\ldots\\
    \end{bmatrix}
\end{align}
Here, $B_{i}$ has $\mu_2$ at the $(1,i+1)$th location and the rest of the elements are zero. Necessary and sufficient condition for the positive recurrence is also found in \cite{quan-li-queue-blockchain} using the analysis of \cite{neuts-mat-geo-sol,quan-li-book}.

\begin{theorem}\label{thm::steady-state-quan}
    {\normalfont \textbf{(Li-Ma-Chang\cite{quan-li-queue-blockchain})}} The Markov process $Q$ of $GI/M/1$ type is positive recurrent iff
    \begin{align}
        \lambda<\frac{b\mu_1\mu_2}{\mu_1+\mu_2}.
    \end{align}
\end{theorem}

This result can be intuitively interpreted as average time to receive $b$ transactions should be more than the sum of the average times of the block-generation process and the blockchain-building process, i.e., $\frac{1}{\mu_1}+\frac{1}{\mu_2}<\frac{b}{\lambda}$. In that sense, one can draw similarities between the result and \cite[Lemma~8]{Sompolinsky2015SecureHT}. Let us denote the maximum $\lambda$ sustainable for $\mathbbm{B}(\alpha,b,\mu_1,\mu_2,\lambda,k')$ as $\lambda(\alpha,b,\mu_1,\mu_2)$.

\begin{proposition} \label{prop::stab-queue-no-adv}
    $\frac{b\mu_1\mu_2}{\mu_1+\mu_2}>\lambda(1,b,\mu_1,\mu_2)>\frac{b\mu_1\mu_2}{\mu_1+\mu_2}-\epsilon$, for any positive $\epsilon$.
\end{proposition}

\begin{Proof}
    Let us call the chain formed from the genesis block until a block $b_h$ as ``the chain associated with block $b_h$.'' The first miner of $\mathbbm{B}(1,b,\mu_1,\mu_2,\lambda,k')$ who builds a block $b_1$ on top of the genesis block at time $t_1$ has the least amount of transactions compared to other blocks on the same height. Now, miner node $j$ experiences a network delay of $t_{1j}\in [0,\Delta_1]$ before receiving $b_1$. This delay essentially hampers the process of transactions joining the blockchain as new blocks will create forks during this delay interval instead of building on top of each other. Thus, to consider the maximum $\lambda$ sustainable in the worst case scenario, we need to take $t_{1j}=\Delta_1$, which makes sure that all blocks mined during $[t_{1},t_{1}+\Delta_1]$ are on height $1$. After $t_1+\Delta_1$, all honest miners will mine a block on height $2$. Clearly, the worst case scenario is mining on top of the first mined block on height $1$, i.e., $b_1$ since it contains the least amount of transactions compared to other blocks on the same height. Similarly, the chain associated with the first jumper mined on height $2$, i.e., $b_2$, which is mined on top of $b_1$, has the least amount of transactions compared to other chains associated with any block on the same height. Following this reasoning, the chain consisting of the first jumper on each height always contains the least amount of transactions, hence the worst case scenario. Finally, note that the content of the jumper chain associated with $\mathbbm{B}(1,b,\mu_1,\mu_2,\lambda,k')$ is the same as the content of the chain formed by the Markovian queue process of \cite{quan-li-queue-blockchain}, which assumes no block is mined during $[t_{i},t_{i}+\Delta_i]$. This implies $\mathbbm{B}(1,b,\mu_1,\mu_2,\lambda,k')$ has the same condition as in Theorem \ref{thm::steady-state-quan}. Note that $k'$ does not really matter as $\alpha=1$.
\end{Proof}

\subsection{Blockchain Design}
In Proposition~\ref{prop::relationship_mu_s}, we established that the relationship between $\mu_1$ and $\mu_2$ has to be kept constant to keep $\overline{p}(\alpha,\mu_1,\mu_2,k)$ at a certain level and $\kappa$ is a safety parameter. Hence, from now on, we assume $\mu_2$ was chosen as $\overline{\kappa}\mu_1$ at the design stage to keep the security at a certain level. As a result, the stability condition in Theorem~\ref{thm::steady-state-quan} and Proposition~\ref{prop::stab-queue-no-adv}, i.e., under no adversarial attack on queue process, can be re-expressed as,
\begin{align}
    \lambda_0 &<bg(b)\frac{f(\alpha,k,\overline{p})}{(1+f(\alpha,k,\overline{p}))}\\
    &=bg(b)\frac{\overline{\kappa}}{(1+\overline{\kappa})}.
\end{align}
If we simply assume that the average delay is linear in block size \cite{linear-prop-delay}, i.e.,
\begin{align}
    \frac{1}{\mu_1}&=\frac{b}{c}+\delta_0, \label{eq::lin-delay-model}
\end{align}
where $c$ is the network speed (measured as transaction/second), the stability condition can be expressed in terms of the security and network parameters as,
\begin{align}
    \lambda_0 &<\frac{bc}{b+\delta_{0}c}\frac{\overline{\kappa}}{1+\overline{\kappa}}. \label{eq::ultimate-design}
\end{align}

The relations above give a blockchain designer a means to analyze critical blockchain parameters such as $\kappa, \alpha, k,b$ and their implications on sustainability and security. In other words, once a desired security level is determined and network conditions are known, the maximum sustainable transaction arrival rate for a blockchain can be obtained using \eqref{eq::ultimate-design}.

Note that the way we derive this equation is by assuming that $\mu_2$ was chosen as $\overline{\kappa}\mu_1$ at the design stage to keep security at a certain level. However, this condition does not take into account any adversarial behavior on the queue process that might try to hamper the progress of the blockchain system. Next, we discuss two different attacks and their effects on sustainability. In general, there are different adversarial attacks that can hamper the growth of a blockchain, and their effects on the model described in Markovian queue process of \cite{quan-li-queue-blockchain} can be analyzed thoroughly. As our goal is to give an insight on how attacks could effect sustainability, we briefly mention two types of attacks.

Every honest block in the longest chain contains $\min(b,|Q_i(t)|)$ transactions, where $Q_i(t)$ is the mempool at the mining time $t$ of that block observed by its honest miner $i$. Hence, every honest block, by assumption, contains the maximum number of transactions possible observed by its honest miner $i$ for its associated chain and does not hamper the queue process. Adversary, on the other hand, by deviating from the protocol, can simply mine empty blocks and publish them immediately as honest miners, which hampers the queue service process.

Under such an attack, even though blocks are mined with rate $\mu_2$ according to the block generation process, the effective mining power drops to $\mu_2\alpha$ since only $\alpha$ fraction of the blocks are non-empty. Hence, the sustainable transaction rate drops to $\lambda_1$, with
\begin{align}
    \lambda_1&=\frac{b\alpha\overline{\kappa}\mu_1^2}{(1+\alpha\overline{\kappa})\mu_1}\\
    &=bg(b)\frac{\alpha\overline{\kappa}}{1+\alpha\overline{\kappa}}. \label{eq::private-attack-sustainable}
\end{align}

Notice, we consider such an attack for purely theoretical reasons in order to understand the theoretical effect of the adversarial attacks on queue-service process since the attack will result in adversarial revenue loss from the transaction fees. Here, as adversary is slowing down by decreasing the fraction of non-empty blocks in the longest chain in the long-run, we can consider attacks that further decrease this fraction. In fact, by slightly changing the private attack strategy and publishing the private chain with empty blocks whenever the lead satisfies some condition, adversary can significantly reduce the service rate of the queue, which we describe next.
\subsection{Sustainable Transaction Rate Under Selfish-Mining}

Here, we describe a stronger attack to hamper the queue-service, which is essentially a selfish-mining attack specifically designed to increase the fraction of the adversarial blocks in the longest chain for the exponential delay model of this paper, where ties are broken in adversary's favor. The attack described here is the optimal selfish-mining strategy for the adversarial model considered in this paper. In other words, the attack results in maximum fraction of adversarial blocks in the longest chain, hence, it is also the worst-case scenario in terms of hampering the queue-service process if the adversarial blocks are empty.

Let $h_u$ denote the length of the longest public chain and $h_v$ denote the length of the longest private chain where $h_v\geq h_u$. Let us call ``honest mining time'' as the time an honest block is mined and ``honest publishing time'' as the time an honest block is shared with all other honest nodes where the adversary delays all honest block publications with maximum allowed delay, i.e., $t_{hi}=\Delta_h$ for all $h$, $i$. We only consider the jumper chain which has the least amount of transactions and is the worst-case scenario, i.e., the honest blocks mined during $[t_{h},t_{h}+\Delta_h]$ are ignored as they all are on the same height $h$ (i.e., they are forked, not rigged).  Consider an adversarial strategy, which we call \emph{queue-service attack}, where adversary mines a private chain with empty blocks as follows: 

\begin{enumerate}
    \item \label{enum::state-1} After an honest mining time, $t_{h_u}$, if there is no private chain, adversary tries to mine on the same height as this lastly mined honest block. If a valid nonce is found at time $t\geq t_{h_u}$ by the adversary, the adversary publishes its block at $\max(t,t_{h_u}+\Delta_{h_u})$ in order to nullify that honest block and starts a private chain on top of its recently published block.    
    \item \label{enum::state-0} At every honest publishing time, $t_{h_u}+\Delta_{h_u}$, if $h_v\geq h_u$, adversary publishes the part of its private chain until (including) the block on height $h_u$. As the ties are broken in adversary's favor and adversarial blocks do not suffer network delay, the adversarial block will be favored by everyone. Then, adversary continues mining its private chain.
\end{enumerate}

The algorithm above is a simple variation of the selfish-mining attack described in \cite{selfish-mining}, where we slightly change the algorithm since adversary controls the network delay and ties are broken in its favor. Although the original goal of selfish-mining attack is to increase the revenues, here we assume that the adversarial blocks are empty in order to hamper the service process of the queue system, i.e., the adversary forgoes the revenues of transaction fees. However, in real crypto systems such as Bitcoin, transaction fees are minimal ($\approx1\%$) compared to coinbase rewards from the blocks hence the attack here is still profitable compared to honest mining.

The resulting state machine has states $\{-1,0,1,2,\ldots\}$. In this sense, state $i$ refers to $h_v-h_u$ at an honest publishing time or an adversarial mining time. Since adversary is delaying every honest block by the maximum allowed delay, the resulting state machine transitions have the same structure as the birth-death process in \eqref{eq::pmatrix},
\begin{align}
    P'=\begin{bmatrix}
        \alpha \sigma' & \alpha\sigma'\rho'+\beta & \alpha\sigma'\rho'^2  & \alpha\sigma'\rho'^3 & \alpha\sigma'\rho'^4 & \ldots\\
        \alpha \sigma' & \alpha\sigma'\rho' & \alpha\sigma'\rho'^2+\beta  & \alpha\sigma'\rho'^3 & \alpha\sigma'\rho'^4 & \ldots\\
        0          & \alpha \sigma' & \alpha\sigma'\rho' & \alpha\sigma'\rho'^2+\beta  & \alpha\sigma'\rho'^3 & \ldots\\
        0          & 0 & \alpha \sigma' & \alpha\sigma'\rho' & \alpha\sigma'\rho'^2+\beta   &  \ldots\\
        \vdots & \vdots & \vdots & \vdots & \vdots & \ddots \\
    \end{bmatrix}. \label{eq::primematrix}
\end{align}

The reasoning is as follows: Assume that there is no private chain and adversary tries to mine on the same height $h_u$ as the last block on the public chain which is an honest block, i.e., the chain is at state $-1$. With probability $\beta$, adversary mines a block on the same height $h_u$, publishes it and nullifies the honest block and we move to state $0$. With probability $\alpha\sigma'\rho'^i$, another honest block is mined on $h_u+1$ followed by $i$ adversarial blocks that are mined during the delay time of this honest block. The first of these adversarial blocks is mined on height $h_u+1$ nullifying the honest block on height $h_u+1$ (the honest block on height $h_u$ stays in the longest chain), the rest ($i-1$ adversarial blocks) are kept private, thus, we transition to state $i-1$. When the chain is at state $0$, i.e., $h_u=h_v$, if an adversarial block is mined (with probability $\beta$), the chain moves to state $1$, where adversary has a private block waiting to be published when the next honest block is mined and published. However, when the chain is at state $0$, if an honest block is mined followed by no adversarial block during its delay which happens with probability $\alpha\sigma'$, we move to state $-1$ from state $0$. If that honest block was followed by $i$ adversarial blocks during its delay, the first one would be published at the end of the honest delay nullifying the honest block and the rest would be kept private, resulting in a transition to state $i-1$. The rest of the cases are similar.

Here, we are interested in the fraction of honest blocks on the longest chain in the long-run. Since $P'$ has the same structure as $P$, using Lemma~\ref{lemma::ramaswami},  we get $\pi'_{-1}=3-\sigma'^{-1}-\alpha^{-1}$ for the steady state distribution of $P'$. Next, using $\pi'_{-1}$, and publishing strategy described earlier, we find the honest fraction of the blocks in the longest chain in the long-run.

\begin{lemma}\label{lemma::queue-attack-honest-prop}
    Under a queue-service attack, $\alpha'=\frac{\pi'_{-1}}{1-\pi'_{-1}\rho'}$ fraction of the blocks in the longest chain are honest in the long-run.
\end{lemma}

\begin{Proof}
    Let $h_u$ denote the length of longest public chain and $h_v$ denote the length of the longest private chain as usual. We calculate the honest fraction by considering the blocks which are published and stay in the longest chain indefinitely using the steady state and transition frequencies. Note that whenever the chain is at state $-1$ and another honest block is mined, the honest block on height $h_u$ stays in the longest chain due to the queue-service attack strategy described earlier,
    \begin{align}
        r_{honest}=\pi'_{-1}\alpha.
    \end{align}
    On the other hand, whenever the chain is at state $-1$ and an adversarial block is mined, it is published and adopted by everyone (the honest block is nullified). Further, whenever the chain is at state $0$ and an honest block is mined followed by $i>0$ adversarial blocks during its delay, the first adversarial block on the private chain is published and adopted by everyone. Lastly, whenever the chain is at state $i>0$ and an honest block is mined, the first adversarial block on the private chain is published and adopted by everyone. Hence,
    \begin{align}    
    r_{adv}&=\pi'_{-1}\beta+\pi'_0\alpha\rho+\sum_{i>0}\pi'_{i}\alpha.
    \end{align}
    
    Using $\pi'_{-1}=(\pi'_{-1}+\pi'_{0})\alpha\sigma'$ and $\sum_{i>0}\pi'_{i}=(1-\pi'_{-1}-\pi'_{0})$, we obtain,
    \begin{align}    
    \alpha'&=\frac{r_{honest}}{r_{adv}+r_{honest}}\\
    &=\frac{\pi'_{-1}}{1-\pi'_{-1}\rho'},
    \end{align}
    completing the proof.
\end{Proof}

Note that, with this strategy, all adversary is doing is to publish some of its empty private blocks whenever honest blocks are published in order to nullify them and stall the queue service. On the other hand, no adversarial block is nullified by honest blocks. Thus, since every adversarial block nullifies an honest block, the attack described above is the optimal selfish-mining attack for our system model \cite[Appendix~B, Proposition~1]{optimal-selfish}. When there is no network delay, i.e., $\kappa=0$, $\alpha'=\frac{\alpha-\beta}{\alpha}$, which agrees with the upper bound of the adversarial relative revenue in selfish-mining when $\gamma=1$, i.e., when ties are broken in the adversary's favor \cite{optimal-selfish}. 

The block generation process of the longest chain has the rate $\mu_2$. The resulting chain, on the other hand, is formed by honest blocks and empty adversarial blocks that cancel some other honest blocks, and it is slightly different from Markovian queue process of \cite{quan-li-queue-blockchain}. However, in the long-run, we know that the model follows the same block generation process with rate $\mu_2$, and $\alpha'$ fraction of the mined blocks remain on the longest chain and the rest are replaced by empty adversarial blocks. Thus, we can consider the maximum $\lambda$ sustainable under the queue-service attack as follows.

\begin{proposition} \label{prop::stab-queue-service-attack}
    $\lambda(\alpha,b,\mu_1,\mu_2)=\lambda(1,b,\mu_1,\mu_2\alpha')$.
\end{proposition}

As a result, the sustainable transaction rate under the worst case queue-service attack can be re-expressed as,
\begin{align}
\lambda_2 = bg(b)\frac{\alpha'\overline{\kappa}}{(1+\alpha'\overline{\kappa})}.\label{eq::queue-attack-sust-rate}
\end{align}
Notice here that, we assume all ties are broken in adversary's favor and adversarial blocks do not suffer any network delay, the results above are extremely pessimistic. It is also worth noting that, in \eqref{eq::queue-attack-sust-rate}, $\overline{\kappa}$ is found by considering worst-case scenario of a double-spending attack where the adversary is assumed to be extremely powerful. $\alpha'$ on the other hand, is found by considering worst-case scenario of a selfish-mining attack, with empty adversarial blocks. Further, it is hard to imagine that the adversary can optimally do both selfish and private attack at the same time to achieve a hampered rate of transactions given in \eqref{eq::queue-attack-sust-rate}. However, as the designer does not know which attack adversary might employ, the rate given in \eqref{eq::queue-attack-sust-rate} is a conservative but safe choice. Moreover, the assumption of the adversary publishing empty blocks considered in this section is due to purely theoretical reasons rather than a realistic situation, since we want the adversary to hamper the queue process as much as it can.

\begin{figure}[t]
	\centerline{\includegraphics[width=0.8\columnwidth]{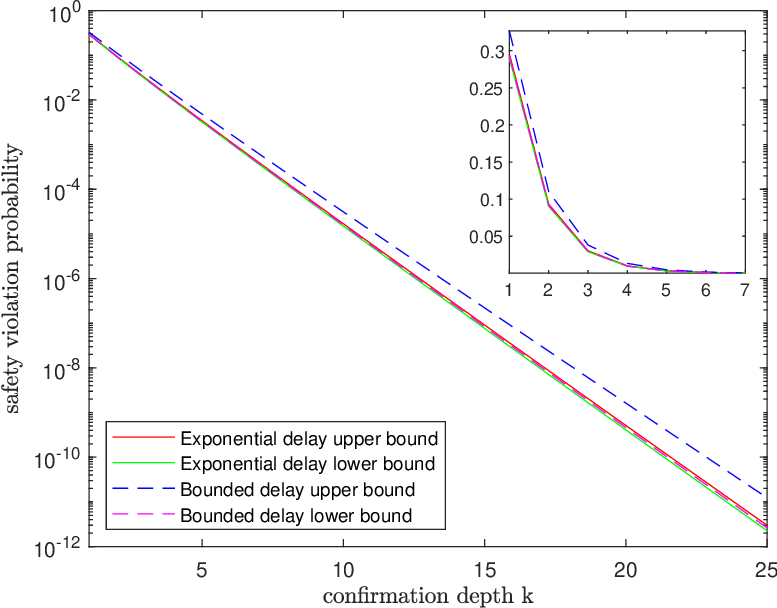}}
	\caption{Safety violation for exponential and bounded delay, $\alpha=0.90$.}
	\label{fig::90sexpvsbndd}
\end{figure}

\begin{figure}[t]
	\centerline{\includegraphics[width=0.8\columnwidth]{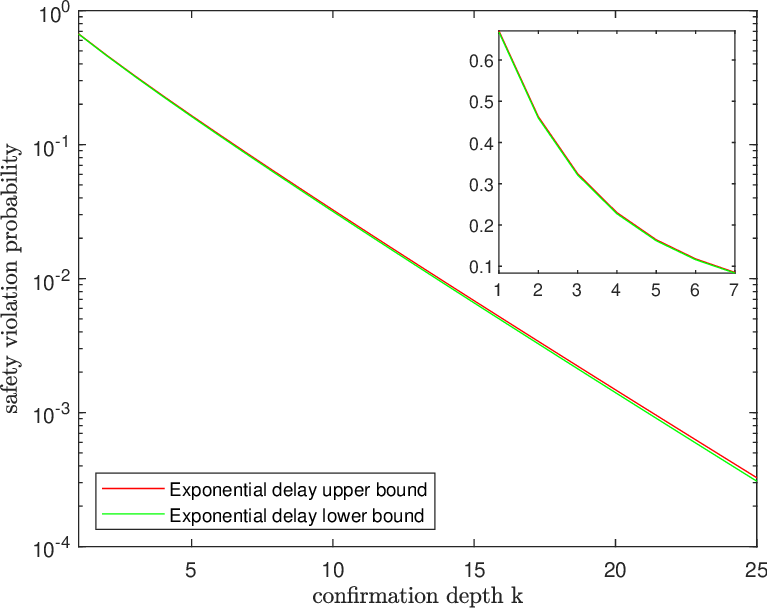}}
	\caption{Safety violation for exponential and bounded delay, $\alpha=0.75$.}
	\label{fig::75expseclat}
\end{figure}

\section{Numerical Results}
We first compare the security-latency results of our model with that of the existing bounded delay model in the literature \cite{our-sec-lat-extended}. We present the results for Bitcoin (BTC) parameters, where we choose $\mu_2=1/600$, $\Delta=10$ for bounded delay, and $\mu_1=\frac{\ln{10}}{4}$ as the $90$th percentile delay for block propagation is around $4$ seconds \cite{DSN-Bitcoin-Monitoring}. \figref{fig::90sexpvsbndd} displays the comparison of two models for $\alpha=0.9$, where it is clear that using an exponential model gives tighter results for upper and lower bounds. The results are not surprising as the bounds in \cite{our-sec-lat-extended} are derived by taking each honest propagation delay as $\Delta=10$ seconds (maximum possible), whereas we take $\Delta_h$ (maximum possible) which is less than $4$ seconds $90$ percent of the time. Note that, if we were to pick $\mu_1=\frac{1}{\Delta}$, where $\Delta$ represents the maximum constant delay in bounded delay models, the resulting security bounds of the exponential model would overlap with those of the bounded delay model, hence, both models agree with each other in that sense, when the safety parameters (fork rates) are the same. However, the exponential delay model with percentiles captures the real propagation delay data of \cite{DSN-Bitcoin-Monitoring} better than the bounded delay model. In \figref{fig::75expseclat}, we present the safety violation probability results for $\alpha=0.75$.  

\begin{figure}
     \centering
     \begin{subfigure}[b]{0.48\textwidth}
         \centering
         \includegraphics[width=\textwidth]{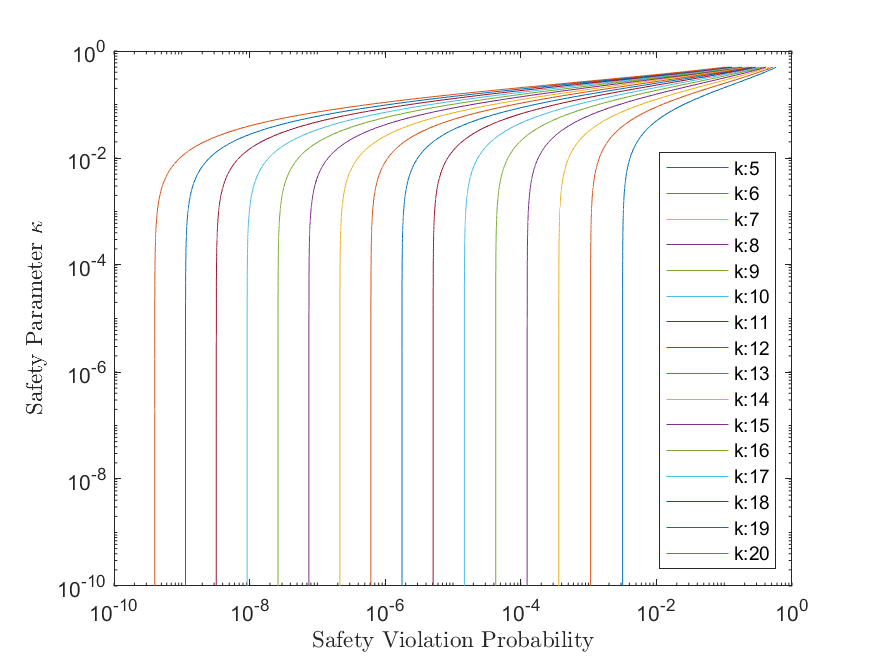}
         \caption{$\overline{p}$ vs $\kappa$}
         \label{fig::pkappa}
     \end{subfigure}
     \hfill
     \begin{subfigure}[b]{0.48\textwidth}
         \centering
         \includegraphics[width=\textwidth]{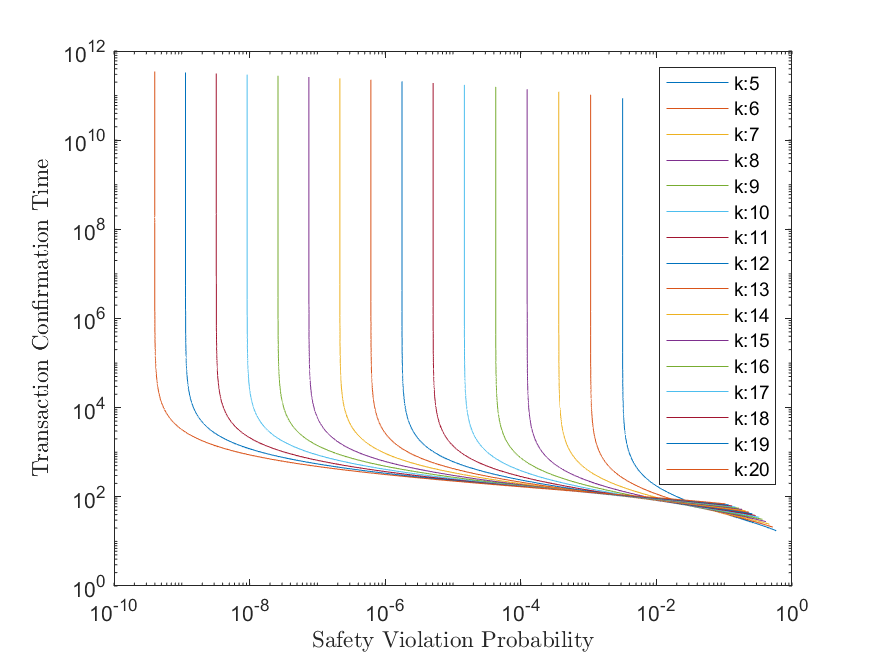}
         \caption{$\overline{p}$ vs $E[C_f]$}
         \label{fig::pconf}
     \end{subfigure}
     \hfill
     \begin{subfigure}[b]{0.48\textwidth}
         \centering
         \includegraphics[width=\textwidth]{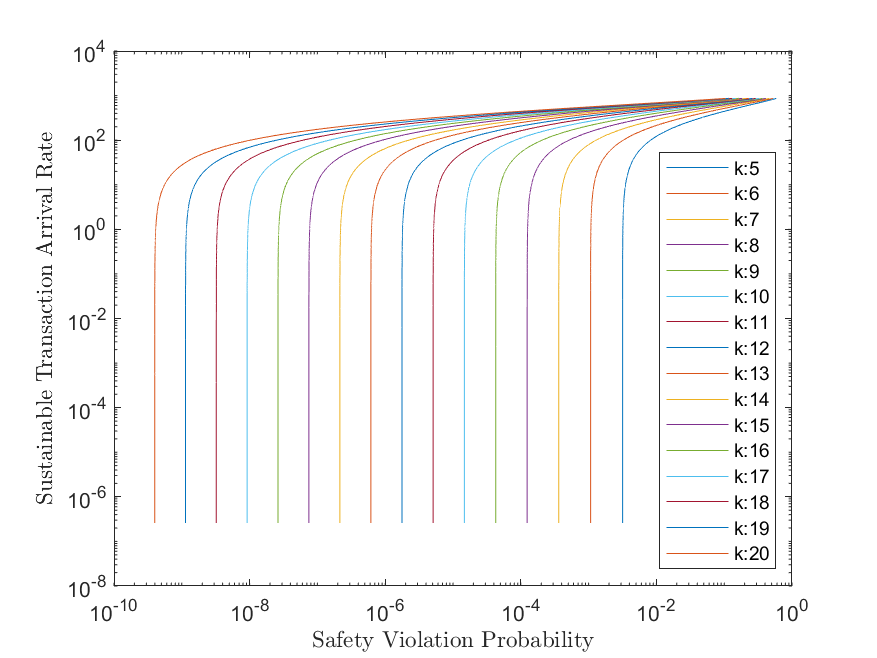}
         \caption{$\overline{p}$ vs $\lambda$}
         \label{fig::prate}
     \end{subfigure}
        \caption{Safety violation probability effect on different parameters, $\alpha=0.9$.}
        \label{fig::three graphs}
\end{figure}

\begin{table}[t]
\begin{center}\footnotesize
\begin{tabular}{|c|c|c|c|c|c|c|c|}
\hline
  Mining Rate & Fault Tolerance&Throughput &Conf. Rule& \multicolumn{3}{c|}{Safety Violation Probability, $\overline{p}$} & Latency  \\
  $\mu_2$ & $\beta_{max}$& $\lambda_0$ &$k$ & $\beta=0.1$&$\beta=0.25$&$\beta=0.4$& $\frac{k}{\mu_2}$  \\ [0.5ex] 
 \hline
 \multirow{3}{*}{$\frac{1}{600s}$ }& \multirow{3}{*}{$0.4993$ }& \multirow{3}{*}{$7.48$ } &$6$ &$0.0012$&$0.1179$&$0.6925$&$3600s$\\
 & &&$22$ &$6.5\cdot10^{-11}$&$8\cdot10^{-4}$&$0.3004$&$13200s$\\
   & &&$149$ &$\leq10^{-14}$&$\leq10^{-14}$&$10^{-3}$&$89400s$\\
 \hline
  \multirow{3}{*}{$\frac{1}{90s}$ }& \multirow{3}{*}{$0.4951$ } & \multirow{3}{*}{$49.05$ }&$7$ &$7\cdot10^{-4}$&$0.0984$&$0.6825$&$630s$\\
 & &&$23$ &$1.2\cdot10^{-10}$&$9\cdot10^{-4}$&$0.3212$&$2070s$\\
   & &&$166$ &$\leq10^{-14}$&$\leq10^{-14}$&$10^{-3}$&$14940s$ \\
 \hline
   \multirow{3}{*}{$\frac{1}{60s}$} & \multirow{3}{*}{$0.4927$ }& \multirow{3}{*}{$72.89$ } &$7$ &$10^{-3}$&$0.1070$&$0.6986$&$420s$\\
 & &&$24$ &$1.2\cdot10^{-10}$&$9\cdot10^{-4}$&$0.3289$&$1440s$\\
   & &&$177$ &$\leq10^{-14}$&$\leq10^{-14}$&$10^{-3}$&$10620s$ 
 \\
 \hline
   \multirow{3}{*}{$\frac{1}{30s}$} & \multirow{3}{*}{$0.4851$ }& \multirow{3}{*}{$141.79$ } &$8$ &$10^{-3}$&$0.1044$&$0.7175$&$240s$\\
 & &&$27$ &$1.9\cdot10^{-10}$&$9\cdot10^{-4}$&$0.3600$&$810s$\\
   & &&$217$ &$\leq10^{-14}$&$\leq10^{-14}$&$10^{-3}$&$6510s$ 
 \\[1ex] 
 \hline
\end{tabular}
\caption{Transaction Capacity, Security and Latency for fixed block size $b$.}
\label{table::safety_table}
\end{center}
\end{table}

\begin{table}[t]
\begin{center}\footnotesize
\begin{tabular}{|c|c|c|c|c|}
\hline
  Mining Rate &Throughput& Adversarial Presence & \multicolumn{2}{c|}{Queue attack}\\
  $\mu_2$ &$\lambda_0$& $\beta$ & $\lambda_1$ & $\lambda_2$ \\ 
 \hline
  \multirow{3}{*}{$\frac{1}{600s}$ }& \multirow{3}{*}{$7.48$ } &$0.1$ &$6.73$&$6.65$\\
 & &$0.25$ &$5.61$&$4.99$\\
  & &$0.4$ &$4.49$&$2.49$ \\
  \hline
  \multirow{3}{*}{$\frac{1}{90s}$ }& \multirow{3}{*}{$49.05$ } &$0.1$ &$44.23$&$43.67$\\
 & &$0.25$ &$36.96$&$32.78$\\
  & &$0.4$ &$29.66$&$16.21$ \\
  \hline
  \multirow{3}{*}{$\frac{1}{60s}$ }& \multirow{3}{*}{$72.89$ } &$0.1$ &$65.79$&$64.95$\\
 & &$0.25$ &$55.05$&$48.76$\\
  & &$0.4$ &$44.23$&$24.00$ \\
  \hline
  \multirow{3}{*}{$\frac{1}{30s}$ }& \multirow{3}{*}{$141.79$ } &$0.1$ &$128.31$&$126.63$\\
 & &$0.25$ &$107.82$&$95.12$\\
  & &$0.4$ &$86.98$&$46.02$ \\
  [1ex] 
 \hline
\end{tabular}
\end{center}
\caption{Throughput decrease under queue-service attacks}
\label{table::throughput_decrease}
\end{table}

Next, in \figref{fig::three graphs}, we present the relationships of safety violation threshold $\overline{p}$ with confirmation rule $k$, safety parameter $\kappa$, expected transaction confirmation time $E[C_f]=\frac{k}{\mu_2}$ and maximum sustainable transaction arrival rate $\lambda$ (under no queue attack) for $\alpha=0.9$ where we choose $b=4500$ (corresponds to approximately $1$ MB of BTC block \cite{btc-tx-per-block}) and $\mu_1=\frac{\ln{10}}{4}$. We note that the shape of the curves in \figref{fig::three graphs} implies that for a given $k$-block confirmation rule, block size $b$ and network delay rate $\mu_1$, one gains almost nothing in terms of safety violation probability by decreasing safety parameter $\kappa$ (and hence the mining rate) in the region where graphs are vertical whereas sustainable transaction rate decreases rapidly and the transaction confirmation time increases significantly. As a result, it is more sensible to choose parameters in the region where the graph in \figref{fig::pkappa} stops its steep rise. Moreover, increasing the $k$-block confirmation rule for a given security threshold allows even more gains in terms of the sustainable transactions per second until a point where we have to trade-off this gain with transaction confirmation time, i.e., more latency. This phenomenon is observable in \figref{fig::pconf} where the tail of the graphs cross each other in the right bottom corner. Further, increasing adversarial presence $\beta$, i.e., lower $\alpha$, shifts the curves rightwards, i.e., increases the safety violation probability, but the shape of the curves remain the same. Hence, the observations we make on \figref{fig::three graphs} apply to general $\alpha$ rather than just $\alpha=0.9$.

Next, as a case study, we consider how the current Bitcoin parameters can be adjusted to achieve better confirmation latency and throughput, while keeping the security guarantees around the same level as it is now. Currently, under the $6$-block confirmation rule and the current mining rate of BTC, i.e., $\kappa=\frac{4}{600\ln{10}}$, with $\beta=0.1$ adversarial presence, we get a safety violation upper bound of $\overline{p}=0.0012$, and a maximum sustainable transaction rate of $\lambda<7.48$, when there is no attack on the queue. Note that under current $6$-block confirmation rule and mining rate, the confirmation latency in Bitcoin is one hour ($3600$ seconds). 

We first consider moving slightly upwards in the curve of \figref{fig::pkappa}, while keeping the safety violation probability below the threshold of $10^{-3}$ for $\alpha=0.9$.
Jumping over from $k=6$ to the curve of $k=7$ and increasing $\kappa$ such that mining rate becomes $\frac{1}{90s}$, i.e., the average block inter arrival time is $90s$, satisfies the threshold. The resulting transaction capacity, security and confirmation latency are reported in Table~\ref{table::safety_table}. Notice that, transaction capacity is increased almost sevenfold from $\approx7$ to $\approx49$ while we also get better confirmation latency of $630$ seconds. The upper bounds on safety violation probability for different adversarial presence of $\beta=0.1$, $\beta=0.25$ and $\beta=0.4$ are reported as well, where we consider the confirmation rules that achieve safety violation probability below $10^{-3}$ for each adversarial presence. For example, with the current Bitcoin parameters, under adversarial presence of $\beta=0.25$, $k=22$ achieves a safety guarantee of $10^{-3}$ whereas when mining rate is increased to $\frac{1}{90}$ seconds, $k=23$ is needed. Still, the new confirmation latency in time units is superior and drops from $\approx3.7$ hours to $\approx0.6$ hours.  

Further moving upwards and jumping over the curves provides even better throughput and confirmation latencies which are also reported in Table~\ref{table::safety_table}. For example, transaction capacities of $\approx140$ with confirmation latencies of a couple minutes are possible with $30$ seconds of block interarrival time. Note that, further increasing mining rate might result in congestion in the network, hence, we do not consider further increases in the mining rate. With the new parameters, the only downside is a slightly decreased fault tolerance. Note that, the actual ultimate fault tolerance for our system model is slightly higher and provided in Appendix~\ref{sec::app-c}. However, in Table~\ref{table::safety_table}, we report the fault tolerance proven in Corollary~\ref{cor::fault-tolerance}, since the result in Appendix~\ref{sec::app-c} is not proven and does not apply to security thresholds obtained from rigged model.

In Table~\ref{table::throughput_decrease}, we consider the throughput decrease under the queue service attacks analyzed in Section~\ref{sec::queue} for the sake of completeness. Essentially, with $\beta=0.1$, the throughput decreases from $\lambda_0$ to $\lambda_1$ (around $\approx10\%$) when the adversary publishes empty blocks. Employing the same strategy together with selfish-mining attack achieves a throughput decrease from $\lambda_0$ to $\lambda_2$ (around $\approx11\%$). When $\beta=0.25$, the percentage of throughput decrease is around $\approx25\%$ from $\lambda_0$ to $\lambda_1$ and around $\approx33\%$ from $\lambda_0$ to $\lambda_2$. When $\beta=0.4$, the percentage of throughput decrease is around $\approx40\%$ from $\lambda_0$ to $\lambda_1$ and around $\approx67\%$ from $\lambda_0$ to $\lambda_2$.

We note that, we do not pose the same optimization problem posed in \cite[Section~IV.B]{cao2023tradeoff} that establishes a trade-off between throughput and latency for a given security level. The optimization problem posed in \cite{cao2023tradeoff} assumes an independent relationship between security threshold $\overline{p}$ and block size $b$ and connects it to the network delay with a linear relationship assumption as in \eqref{eq::lin-delay-model}. In short, authors try to find maximum throughput by optimizing block size, mining rate and confirmation rule under the restriction of a security threshold $\overline{p}$. Thus, it treats the block size an independent quantity compared to the security threshold $\overline{p}$. As we mentioned in Section~\ref{sec::sec-lat analysis}, adversary
can potentially discard more than one transaction (up to $b$ transactions if they happen to be in the same block) with the same attack if successful. Hence, $b$ and $\overline{p}$ are inherently dependent parameters and should not be treated differently. To understand the situation better, consider the following analogous situation. For example, when a transaction involves a major sum of fund transfer between accounts, more confirmations are needed since the security threshold $\overline{p}$ we want is higher when a larger sum of money is at risk. Similarly, when the block size increases, we need a higher security threshold $\overline{p}$, since the total sum of funds transferred in a block becomes higher. Hence, $\overline{p}$ and $b$ are dependent parameters and should be treated as such. For this reason, in our numerical results and considerations for Bitcoin parameters, we considered a fixed block size $b$ and $\overline{p}$.

\section{Conclusion}
In this paper, we developed the security-latency analysis for exponential network delay by considering the rigged jumper chain. We reconstructed the Markov chains and random walk models of the security-latency analysis for exponential network delay \cite{guo-btc-sec-lat, our-sec-lat-extended}, derived their steady state behaviors. By dividing the honest and adversarial race into three phases, we derived the relevant distributions for each phase and combined them to give an upper bound on safety violation probability. We derived fault tolerance conditions under which the upper bound on safety violation probability vanishes. We also showed that the upper bound on safety violation probability is a function of the ratio of the mining rate and the network delay rate, which we called $\kappa$, a safety parameter. 

Using the queue analysis of \cite{quan-li-queue-blockchain}, we re-expressed the sustainable transaction rate condition in terms of block size $b$, $\kappa$ and network parameters. We connected the security-latency analysis to the queue analysis by assuming a designer which first chooses a safety parameter $\kappa$ to achieve a security threshold and then uses this same $\kappa$ to obtain the sustainable transaction rate under no queue attack. We considered two different queue attack strategies that can drop the sustainable transaction rate and analyzed their drop rate. The question remains whether one could devise a single optimal attack in a way that could give an upper bound both on safety violation and queue-service. 

\appendices

\section{What if $tx$ waits in the mempool-queue?}\label{sec::app-b}
In Section~\ref{sec::sec-lat analysis}, we assumed that, at the time of arrival of $tx$, $|Q(\tau)|\leq b$. Here, we give a method that applies to general mempool size at the time of arrival of $tx$.

Adopting the notation of \cite{quan-li-queue-blockchain}, the stationary vector of queue $Q$ is written as,
\begin{align}
    \mathbbm{\pi}_j&=(\pi_{0,j},\pi_{1,j},\ldots,\pi_{b,j}), \quad j\geq0.
\end{align}
A recursive formula for finding $\pi=(\pi_0,\pi_1,\ldots)$ is given in \cite[Theorem~2]{quan-li-queue-blockchain}. Note that, the state of the system at time $t$ is taken as $w(t)=(|B(t)|,|Q(t)|)$, where $|B(t)|$ and $|Q(t)|$ are the number of transactions in the block pending to be published and the queue, respectively. Hence, the probability that $tx$ is included in the next $m$th honest block after $\tau$ (denoted as $b_m$) is,
\begin{align}
    P(tx \in b_m)=\sum_{i=0}^{b}\sum_{j=(m-1)b}^{mb-1}\pi_{i,j}.
\end{align}
Note that $\Pi$ is the distribution of the adversarial lead when the first honest block-generation process of the confirmation interval starts. We represent this vector with $\mathbf{d}$ where $d(i)=P(\Pi=i)$. Next, we convert the distribution $C$ derived in Lemma~\ref{lemma::lower-geo} to a Markov chain as follows,
\begin{align}
    P_C=\begin{bmatrix}
        P'_C(0) & P'_C(1) & P'_C(2)  & P'_C(3) & \ldots\\
        P_C(0) &P_C(1) & P_C(2)  & P_C(3) &  \ldots\\
        0          & P_C(0) &P_C(1) & P_C(2)  &  \ldots\\
        0          & 0 & P_C(0) &P_C(1) &   \ldots\\
        0          & 0 & 0 & P_C(0) &   \ldots\\
        \vdots & \vdots & \vdots  & \vdots & \ddots \\
    \end{bmatrix},
\end{align}
where $P'_C(i)=P_C(i+1)+\alpha\sigma(\rho^i-\rho^{i+1})$. $P_C$ is the transition of a Markov chain where the $(i,j)$th element represents a transition from lead $i$ to $j$ between publications of honest blocks. For $i>0$, this is a simple conversion of Lemma~\ref{lemma::lower-geo}. For $i=0$, the elements are adjusted by taking into account whether adversarial blocks are arrived during the block-generation process or not. The adjustment is made according to the observation that if all arrivals are during blockchain-building process, then all adversarial blocks can be built on top of the honest block waiting to be published, hence the shift in some elements.

After $\tau$, $m-1$ blocks are mined and published before $tx$ can enter the chain with $b_m$. Denoting the publication time of $b_{m-1}$ as $\tau'$, the $m$th honest block after $\tau$ is the first honest block after $\tau'$ and $tx$ is going to be included in this block. Further, the lead vector of the adversary transitions to $\mathbf{d}P_C^{m-1}$ right at $\tau'$, thus, we call the new lead at $\tau'$ as $D_m$ where $P(D_m=i)=\mathbf{d}P_C^{m-1}(i)$. From $\tau'$ onwards, repeating the analysis of confirmation interval and post-confirmation race of Section~\ref{sec::sec-lat analysis} gives,

\begin{align}
     p(\alpha,\mu_1,\mu_2,k) &\leq    \overline{p}(\alpha,\mu_1,\mu_2,k)\\
    &=\sum_{m=1}^{\infty}P(tx \in b_m)\cdot P(D_m+S_k+\Pi'\geq k)\\
    &=\sum_{m=1}^{\infty}\left(\sum_{i=0}^{b}\sum_{j=(m-1)b}^{mb-1}\pi_{i,j}\right)\cdot P(D_m+S_k+\Pi'\geq k)
\end{align}
Here, observing numerical evaluations, we note that the longer the $tx$ waits in the mempool queue, the less the adversarial chance to have a high lead. In other words, the lead decreases while the $tx$ waits in the mempool queue, e.g., $P(D_m\geq i)\geq P(D_{m+1}\geq i)$. We believe this is due to the inspection paradox that slightly overestimates the lead at a random $\tau$, but its effect decreases as the $tx$ waits in the mempool queue.
\section{Ultimate Fault Tolerance}\label{sec::app-c}

\begin{observation}
    The ultimate fault tolerance for exponential delay model is,
    \begin{align}
        \beta<\frac{1-\beta}{1+(1-\beta)\kappa}.\label{eq::ult-tol-conj}
    \end{align}
    In other words, any transaction is safe after waiting long enough if \eqref{eq::ult-tol-conj} is satisfied and no transaction is safe if \eqref{eq::ult-tol-conj} is not satisfied.
\end{observation}

We do not provide any proof for the observation above as it is out of the scope for this paper and requires another model to rigorously prove it. The bound in \eqref{eq::ult-tol-conj} is based on the observation that in the long-run, adversarial chain grows with rate $\beta\mu_2$ whereas honest chain grows by $\frac{\alpha\mu_2}{1+\kappa\alpha}$. Alternatively, note that, when adversary delays all honest blocks by maximally allowed time $\Delta_h$ and mines its own private chain (no rigged blocks), the lead is finite if $3-\sigma'^{-1}-\alpha^{-1}>0$. This condition can be re-expressed as $\alpha>\frac{1}{2-\kappa\beta}$ or as in \eqref{eq::ult-tol-conj}.

Equation \eqref{eq::ult-tol-conj} implies that the safety parameter is bounded by,
\begin{align}
     \kappa<\frac{1}{\beta}-\frac{1}{\alpha},
\end{align}
which is the same condition as the ultimate fault tolerance of bounded delay models \cite{nakamoto-always-wins} and \cite[Section~3.3]{cao2023tradeoff}. This implies that the maximum sustainable transaction rate (under no queue-attack) is,
\begin{align}
     \lambda&<bg(b)\frac{\kappa}{1+\kappa}\\
     &<bg(b)\frac{1-2\beta}{1-\beta-\beta^2},
\end{align}
whereas the sustainable transaction rate further drops to,
\begin{align}
     \lambda&<bg(b)\frac{\alpha'\kappa}{1+\alpha'\kappa},
\end{align}
under the worst-case queue-service attack.
 
\bibliographystyle{ieeetr}
\bibliography{blockchain}

\end{document}